\documentclass[twocolumn]{aastex62}

\usepackage{color}\usepackage{graphicx}
\usepackage{amsmath}	
\usepackage{amssymb}	
\usepackage{ulem}
\usepackage[figuresleft]{rotating}
\newcommand{\HI}{H{\,\small I}}

\newcommand{\ltsima} {$\; \buildrel < \over \sim \;$}
\newcommand{\gtsima} {$\; \buildrel > \over \sim \;$}
\newcommand{\lta} {\lower.5ex\hbox{\ltsima}}
\newcommand{\kmsMp}{km s$^{-1}$\,Mpc$^{-1}$}
\newcommand{\gta} {\lower.5ex\hbox{\gtsima}}
\newcommand{\Kkmspc}{K km\,s$^{-1}$ pc$^{2}$}
\newcommand{\kms}{km\,s$^{-1}$}

\newcommand{\lya}{Ly$\alpha$}

\newcommand{\ci}{[C\,{\sc i}]}
\newcommand{\cii}{[C\,{\sc ii}]}

\accepted{ApJ}

\shorttitle{The cold environment of MAMMOTH-I}
\shortauthors{Emonts et al.}

\begin{document}

\title{The cold circumgalactic environment of MAMMOTH-I: dynamically cold gas in the core of an Enormous Ly$\alpha$ Nebula}

\correspondingauthor{Bjorn Emonts}
\email{bemonts@nrao.edu}
\correspondingauthor{Zheng Cai}
\email{zcai@tsinghua.edu.cn}

\author{Bjorn H.\,C. Emonts}
\affiliation{National Radio Astronomy Observatory, 520 Edgemont Road, Charlottesville, VA 22903}

\author{Zheng Cai}
\affiliation{Physics Department and Tsinghua Centre for Astrophysics, Tsinghua University, Beijing 100084, China}
\affiliation{University of California Observatories, Lick Observatory, 1156 High Street, Santa Cruz, California 95064, USA}

\author{J. Xavier Prochaska}
\affiliation{Department of Astronomy and Astrophysics, University of California, 1156 High Street, Santa Cruz, California 95064, USA}
\affiliation{University of California Observatories, Lick Observatory, 1156 High Street, Santa Cruz, California 95064, USA}

\author{Qiong Li}
\affil{Department of Astronomy, School of Physics, Peking University, Beijing 100871, P. R. China}
\affil{Kavli Institute for Astronomy and Astrophysics, Peking University, Beijing, 100871, P. R. China}

\author{Matthew D. Lehnert}
\affiliation{Sorbonne Universit\'{e}, CNRS, UMR 7095, Institut d'Astrophysique de Paris, 98bis bvd Arago, 75014, Paris, France}



\begin{abstract}

The MAMMOTH-I Nebula at redshift 2.3 is one of the largest known Ly$\alpha$ nebulae in the Universe, spanning $\sim$440\,kpc. Enormous \lya\ nebulae like MAMMOTH-I typically trace the densest and most active regions of galaxy formation. Using sensitive low-surface-brightness observations of CO(1-0) with the Very Large Array, we trace the cold molecular gas in the inner 150 kpc of the MAMMOTH-I Nebula. CO is found in four regions that are associated with either galaxies or groups of galaxies that lie inside the nebula. In three of the regions, the CO stretches up to $\sim$30 kpc into the circum-galactic medium (CGM). In the centermost region, the CO has a very low velocity dispersion (FWHM$_{\rm CO}$\,$\sim$\,85\,\kms), indicating that this gas is dynamically cold. This dynamically cold gas coincides with diffuse restframe optical light in the CGM around a central group of galaxies, as discovered with the Hubble Space Telescope. We argue that this likely represents cooling of settled and enriched gas in the center of MAMMOTH-I. This implies that the dynamically cold gas in the CGM, rather than the obscured AGN, marks the core of the potential well of this \lya\ nebula. In total, the CO in the MAMMOTH-I Nebula traces a molecular gas mass of M$_{\rm H2}$\,$\sim$\,1.4\,($\alpha_{\rm CO}$/3.6)\,$\times$\,10$^{11}$\,M$_{\odot}$, with roughly 50$\%$ of the CO(1-0) emission found in the CGM. Our results add to the increasing evidence that extended reservoirs of molecular gas exist in the CGM of massive high-z galaxies and proto-clusters.

\end{abstract}



\section{Introduction}
\label{sec:intro}

Massive galaxies at high redshifts are known to co-evolve with large reservoirs of warm circum-galactic medium (CGM), generally detected in Ly$\alpha$ (T\,$\sim$\,10$^{4}$\,K). Giant nebulae of Ly$\alpha$-emitting gas that extend on scales of $\ga$100\,kpc were first discovered around high-redshift radio galaxies \citep[e.g.,][]{djo87,mcc87,mcc90,mcc95,hec91,eal93,oji96,oji97,vil02,vil03,vil06,vil07,pen01,reu03,reu07,mil06,swi15,ver17}. For a long time, similar giant Ly$\alpha$ nebulae around radio-quiet high-$z$ galaxies remained elusive (see \citealt{ber99} for a rare example). However, more recent studies revealed Enormous Ly$\alpha$ Nebulae (ELANe) stretching across $\sim$300$-$500 kpc around a number of radio-quiet quasi-stellar objects (QSOs) \citep[][]{can14,martin14,hen15,bor16,cai17b,cai18,arr19,cai19}. In fact, Ly$\alpha$ emission on 10s$-$100 kpc scale now appears to be ubiquitously associated with QSOs \citep{bor16,arr19,cai19}. 

The most extreme nebulae of Ly$\alpha$-emitting gas cover regions of several 100 kpc and luminosities of $L\sim$10$^{44-45}$ erg\,s$^{-1}$ \citep[see reviews by][]{mil08,can17}. These ELANe often trace large galaxy overdensities, representing regions of massive galaxy and cluster formation \citep[e.g.,][]{hat09,dan14,hen15,cai17b,arr18a}.

Several mechanisms have been proposed to be responsible for creating the Ly$\alpha$ emission in these giant nebulae, including photo-ionization by strong ultra-violet (UV) radiation \citep{gou96,can05,gea09,kol10,ove13}, resonant scattering of Ly$\alpha$ emission from an embedded source \citep{dij09a,hay11,can14,gea14,gea16}, shock-heating by feedback and outflows \citep{tan00,tan01,wil05}, or cooling radiation from gas that settles deep in the potential well \citep{far01,yan06,dij09b,fau10,ros12}. Observational evidence for gas infall or accretion flows has been suggested for a number of giant Ly$\alpha$ nebulae \citep{vil07,mar15,gul16,ver17,arr18a}.

A question that arises is how do these enormous gas reservoirs evolve, and ultimately drive the evolution of the massive galaxies that are forming within these nebulae? To study the direct connection between these gaseous nebulae and the stellar growth of massive galaxies, it is essential to trace gas at temperatures well beyond the limit of Ly$\alpha$ cooling (i.e, T\,$<<$\,10$^{4}$\,K). Absorption-line studies revealed that the halos of massive QSOs contain cool gas with both a high covering fraction of optically thick \HI\ absorbers ($\sim$60$\%$) and metal enrichment beyond the typical virial radius (r$_{\rm vir}$\,$\sim$\,160\,kpc) \citep[e.g.,][]{hen06,hen07,pro13,pro14,lau16}. In some of the largest ELANe ($\ge$400 kpc), comparing the emission- and absorption-line properties of Ly$\alpha$, C\,IV and He\,II revealed that the cool gas must have a clumpy morphology that likely occupies the high-end tail of the distribution of volume densities, with n\,$\ge$\,1$-$3\,cm$^{-3}$ and cloud sizes $\le$20$-40$\,pc \citep{hen06,hen15,hen07,can14,can18,arr15,arr18a}.

The ultimate reservoir of halo gas that can fuel widespread star-formation is cold molecular gas (T $\sim$ 10$-$100\,K). Widespread reservoirs of cold molecular gas can be efficiently traced with observations of the low-$J$ transitions of carbon-monoxide, CO($J$,$J-1$) \citep[e.g.,][]{pap00,ivi11}. Carbon emission in the form of \ci\ or \cii\ \citep{pap04,car13,bis17} can also be a good tracer of widespread and quiescent reservoirs of cold gas, as \citet{cic15} observed around a quasi-stellar object at $z$\,=\,6.4. These and other studies have added to the growing amount of evidence that widespread molecular gas can be present in the circum-galactic environments of massive high-z galaxies \citep[e.g.,][]{emo14,emo15,emo16,gin17,dan17,fra18,fuj19}.

Direct evidence that cold molecular gas can be spread across many tens of kpc scales within the CGM of giant Ly$\alpha$ nebulae came through observations of the Spiderweb Galaxy at $z$\,=\,2.2 \citep{mil06,emo16}. CO(1-0) observations sensitive to low-surface-brightness emission revealed that the CGM of the Spiderweb contains $\sim$10$^{11}$ M$_{\odot}$ of cold molecular gas spread across $\sim$70 kpc, providing evidence for the existence of a cold, multi-phase CGM across the central parts of the Ly$\alpha$ nebula \citep{emo16}. Subsequent observations of CO(4-3) and \ci\ $^{3}$P$_{1}$-$^{3}$P$_{0}$ revealed that the molecular CGM in the Spiderweb has excitation conditions and carbon abundances similar to the interstellar medium in starforming galaxies \citep{emo18}. This cold molecular medium fuels in-situ star formation detected across the CGM with deep HST imaging \citep{hat08}, providing a direct link between a giant Ly$\alpha$-emitting nebula and the stellar buildup of a massive forming galaxy.

In this paper, we present observations of extended molecular gas across a radio-quiet ELAN, namely the MAMMOTH-I Nebula at $z$\,=\,2.3. The MAMMOTH-I Nebula is one of the most extreme ELANe found to date, with Ly$\alpha$ emission stretching across $\sim$440\,kpc \citep{cai17b}. It is located in the extreme galaxy overdensity BOSS1441, which traces a proto-cluster \citep[][]{cai17a,arr18b,muk19}. An 850$\mu$m continuum source at the center of MAMMOTH-I indicates the presence of a starburst and obscured Active Galactic Nucleus (AGN), with a combined energy sufficient to power the Ly$\alpha$ nebula either through photo-ionization or shock-heated outflows \citep{arr18b}. With the goal of studying studying cold gas that can drive star formation in MAMMOTH-I, we used NSF's Karl G. Jansky Very Large Array (VLA) in its most compact D-configuration to obtain sensitive low-surface-brightness observations of CO(1-0). In Sect.\,\ref{sec:observations} we describe the observations, in Sect.\,\ref{sec:results} we present the results of our study, followed by a discussion in Sect.\,\ref{sec:discussion}.

Throughout this paper, we assume the following cosmological parameters: H$_{0} = 71$\,\kmsMp, $\Omega_\textrm{M} = 0.27$ and $\Omega_{\Lambda} = 0.73$, i.e., 8.3 kpc/$^{\prime\prime}$ and $D_{\rm L} = 18800$ Mpc at z\,=\,2.31 \citep{wri06}.

\section{Observations}
\label{sec:observations}

\subsection{Radio data}
We observed MAMMOTH-I for 14\,hrs on-source with the VLA in D-configuration from February $-$ April 2017 (ID: VLA/17A-174). We used two overlapping sets of 8 spectral windows, containing 128 channels of 1 MHz each, to cover a contiguous bandwidth of $\sim$1\,GHz. We placed this band so that it includes the redshifted CO(1-0) line at $\nu_{\rm obs}$\,$\sim$\,34.808 GHz ($\nu_{\rm rest}$\,=\,115.27 GHz). The observations were done using 3C\,286 for bandpass and flux calibration. The complex gains were corrected with intermittent short scans every $\sim$6.5 minutes on the secondary VLA calibrator J1419+3821, which is located at $\sim$4.5$^{\circ}$ distance from MAMMOTH-I. 

The VLA data were reduced with the Common Astronomy Software Applications (CASA; \citealt{mcm07}; CASA team et al. in prep.). We imaged the data using natural weighting and a spectral resolution of 30 \kms. Because of the faintness of the CO, no deconvolution was applied. The resulting root-mean-square (rms) noise level is 0.057 mJy\,beam$^{-1}$ channel$^{-1}$. The synthesized beam is 2.6$^{\prime\prime}$\,$\times$2.3$^{\prime\prime}$, with a position angle (PA) of $-75.2^{\circ}$. The CO emitters that we discuss in this paper lie within $\sim$10$^{\prime\prime}$ of the pointing center of the observations (RA\,=\,14$^{\rm h}$41$^{\rm m}$24.0$^{\rm s}$; dec\,=\,40$^{\circ}$\,03$^{\prime}$\,10.0$^{\prime\prime}$), hence primary beam corrections are negligible. All velocities in this paper are relative to the CO(1-0) redshift, $z_{\rm CO}$\,=\,2.3116\,$\pm$\,0.0004 (Sect.\,\ref{sec:nature}).

\subsection{Optical data}

We observed MAMMOTH-I with the Hubble Space Telescope (HST) Wide Field Camera 3 (WFC3) in the F160W filter (ID: 14760; PI: Z.\,Cai). For our target at $z$\,=\,2.3116, the infra-red F160W filter covers the restframe optical wavelength range of roughly 4200\,$-$\,5100 \AA. The observations were done on 2017 September 07 with an exposure time of 2212 sec. The data were pipeline calibrated using the data processing software system version COMMON 2017$\_$2a. We then compared our HST image to data from the Gaia Data Release 2 \citep{gai16,gai18} and found two bright stars that appear in both data sets, where they have the same coordinates to within 0.04$^{\prime\prime}$. As we will show in Sect.\,\ref{sec:individual}, the galaxies in regions A and C also align perfectly with the CO(1-0) emission. We therefore did not need to correct the astrometry of the HST data. We note that trying to align the HST image with any of the galaxies in region B (Sect.\,\ref{sec:individual}) would require a shift of $\ga$0.6$^{\prime\prime}$. This would result in offsets between the CO(1-0) and the galaxies in the other three regions (A, C, and D), as well as with the two bright Gaia-catalogue stars in the wider HST field. We therefore assume that the astrometry of our HST data is accurate to $\leq$0.1$^{\prime\prime}$, as typically assumed for pipeline-calibrated HST data. We use these HST data as a reference to the CO(1-0) results. A full description and analysis of the HST data will be provided in a forthcoming paper (Cai et al. in prep).

We also observed MAMMOTH-I with the Keck Cosmic Web Imager (KCWI), which provides integral-field spectroscopy data in the rest-frame ultraviolet. Observations were done on 2018 May 18 with an on-source exposure time of 1 hour. We used the Blue Medium Grating with the Large Slicer, resulting in a spectral resolution of 2000 and field-of-view of 33$^{\prime\prime}$\,$\times$\,20$^{\prime\prime}$. This setup allowed us to simultaneously cover the Ly$\alpha$ and CIV lines. A standard data reduction was performed. Details on the KCWI data will be described in an forthcoming paper (Li et al. in prep.)

\begin{figure*}
\centering
\includegraphics[width=\textwidth]{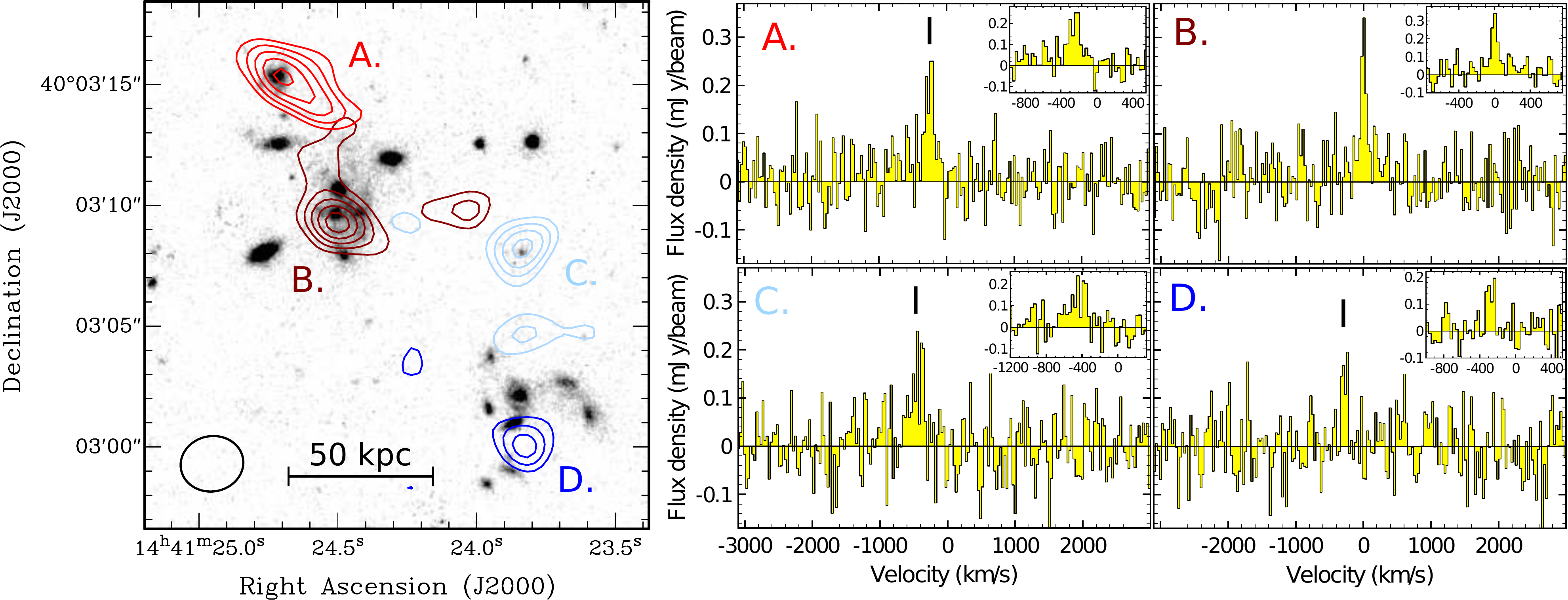}
\caption{CO(1-0) emission in the MAMMOTH-I Nebula. {\sl Left:} total intensity contours of the CO(1-0) emission in four regions overlaid onto the HST/WFC3 160W image from Cai et al. (in prep). Each of the four regions contains a spatially and kinematically distinct CO feature that is visualized with a specific color, as will be further explained with CO channel maps in Fig.\ref{fig:chanmaps}. The total intensity CO emission in each of the four regions was derived by summing the emission across 120 \kms\ (4 channels) and in a box of 100$\times$100 kpc around the peak of the emission. Contour levels: 2.5, 3.5, 4.5, 5.5, 6.5\,$\times$\,$\sigma$, with $\sigma$\,=\,0.0037 Jy\,beam$^{-1}$\,$\times$\,\kms\ the root-mean-square (rms) noise level. {\sl Right:} CO(1-0) spectra taken against the peak emission in each of the four regions. The spectra are centered on the redshift of the CO(1-0) in region B, $z_{\rm CO}$\,=\,2.3116\,$\pm$\,0.0004. The scaling of the axes is the same for all four plots. The insets show a zoom around the detected CO lines.}
\label{fig:mom0}
\end{figure*}

\section{Results}
\label{sec:results}

\begin{figure*}
\centering
\includegraphics[width=0.90\textwidth]{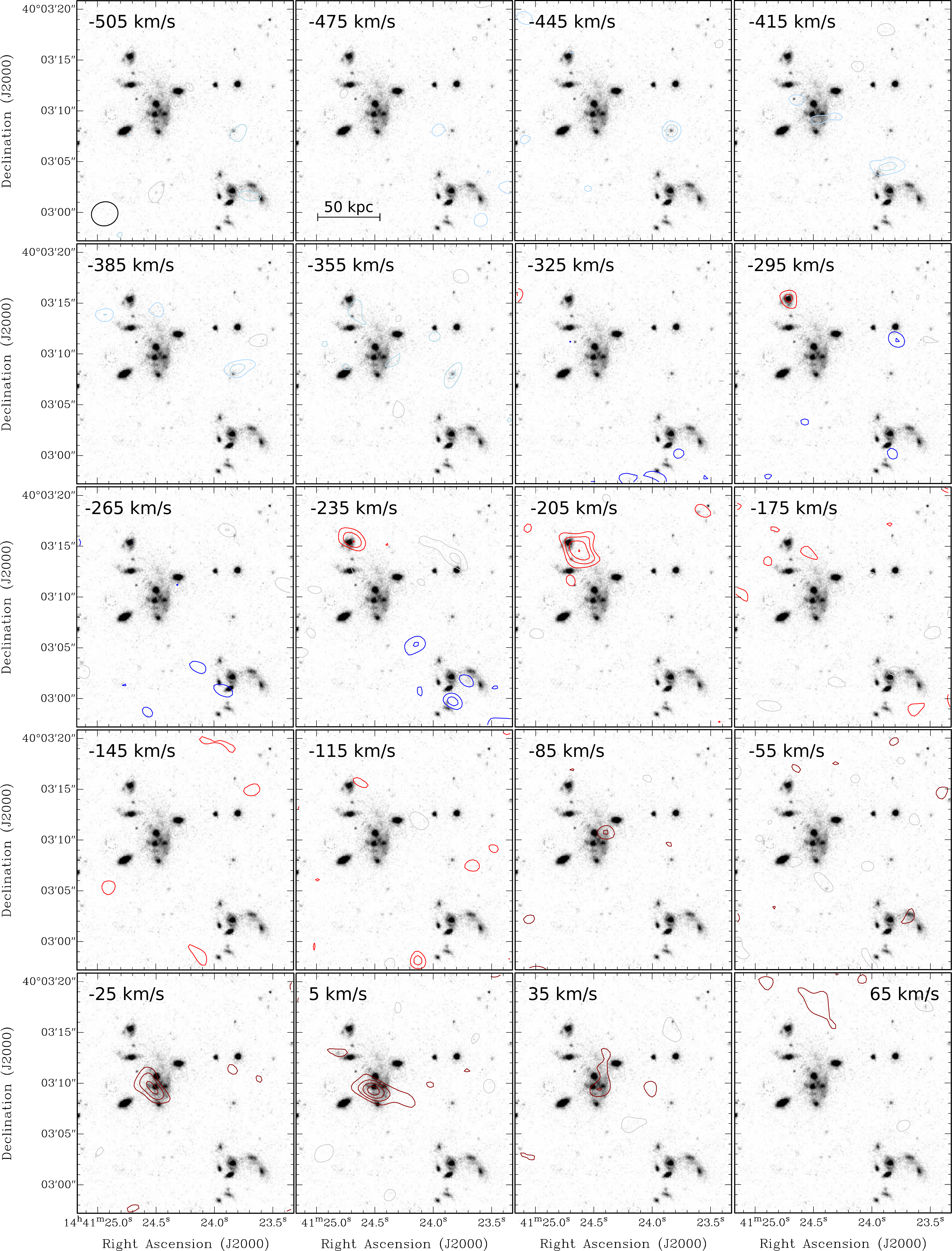}
\caption{Kinematics of the cold molecular gas. The channel maps show contours of the CO(1-0) emission at different velocities overlaid onto the HST/WFC3 F160W image. The colors of the contours match to the colors of the four CO(1-0) features visualized in Fig.\,\ref{fig:mom0}. Contour levels: -3.5, -2.5 (grey), 2.5, 3.5, 4.5, 5.5 (color) $\times$ $\sigma$, with $\sigma$\,=\,0.057 mJy\,beam$^{-1}$ the rms noise level. The synthesized beam of the CO data is show in the bottom-left corner of the first panel.}
\label{fig:chanmaps}
\end{figure*}

We detect CO(1-0) emission of cold molecular gas in the central $\sim$150 kpc region of the MAMMOTH-I Nebula. Figs.\,\ref{fig:mom0} and \ref{fig:chanmaps} show that the CO(1-0) emission is spatially and kinematically associated with four regions, representing galaxies or groups of galaxies. We will first present the CO(1-0) properties of the four region in Sect.\,\ref{sec:individual} and Table \ref{tab:properties}, followed by an overview of the overall CO results for the MAMMOTH-I Nebula in Sect.\,\ref{sec:overview}. 

\begin{deluxetable*}{lcccc}[htb!]
\tablecaption{CO(1-0) properties of MAMMOTH-I. \label{tab:properties}}
\tablecolumns{5}
\tablenum{1}
\tablewidth{0pt}
\tablehead{
\colhead{Region} & {A} & {B} & {C} & {D}
}
\startdata
v$_{\rm CO}$$^{\dagger}$ (\kms) & -245\,$\pm$\,15 & 0 & -445\,$\pm$\,25 & -280\,$\pm$\,15 \\
FWHM$_{\rm CO}$ (\kms) & 170\,$\pm$\,35 & 85\,$\pm$\,20 & 230\,$\pm$\,55 & 105\,$\pm$\,25 \\
$I_{\rm CO}$ (Jy\,bm$^{-1}$\,$\times$\,\kms) & 0.044\,$\pm$\,0.010 & 0.042\,$\pm$\,0.006 & 0.039\,$\pm$\,0.014 & 0.019\,$\pm$\,0.008 \\
$L^{\prime}_{\rm CO}$ (\Kkmspc) & 1.2\,$\pm$\,0.3\,$\times$10$^{10}$ & 1.1\,$\pm$\,0.2\,$\times$10$^{10}$ & 1.0\,$\pm$\,0.4\,$\times$10$^{10}$ & 0.5\,$\pm$\,0.2\,$\times$10$^{10}$\\
\enddata
\tablecomments{The velocity of the CO peak emission, v$_{\rm CO}$, is given with respect to the CO(1-0) redshift of region B, i.e. $z_{\rm CO}$\,=\,2.3116. FWHM$_{\rm CO}$ is the full width at half the maximum intensity of the CO(1-0) profile. Both v$_{\rm CO}$ and FWHM$_{\rm CO}$ were derived by fitting a Gaussian function to the CO line profile taken at the location of the peak of the CO(1-0) emission. The errors in v$_{\rm CO}$ and FWHM$_{\rm CO}$ reflect the errors in the fitting procedure. $I_{\rm CO}$ is the CO(1-0) total intensity, which we derived by integrating the CO(1-0) emission from Fig.\,\ref{fig:mom0} (left) and subsequently correcting this value to reflect the true velocity range that the CO(1-0) emission spans at zero velocity. For the latter, we derive a scaling factor by taking the integrated CO(1-0) intensity of the emission-line profile taken at the location of the peak of the CO emission and dividing this by the CO peak intensity from Fig.\,\ref{fig:mom0} (left). Because Fig.\,\ref{fig:mom0} (left) shows the integrated CO intensity across 4 channels (120 \kms), this correction factor is larger in regions with larger FWHM$_{\rm CO}$, and ranges from 1.1 in region D to 1.9 in region C. $L^{\prime}_{\rm CO}$ is the CO(1-0) luminosity, derived following \citet{sol05}. For $I_{\rm CO}$ and $L^{\prime}_{\rm CO}$, the errors consist of three different errors added in quadrature, namely the error in determining the total intensity CO emission from Fig.\,\ref{fig:mom0}, the error in fitting the emission-line profile against the CO peak, and a statistical error due to the noise in the spectrum (Eqn. 2 of \citealt{emo14}, see also \citealt{sag90}).}
\end{deluxetable*}

\subsection{CO(1-0) in the individual regions}
\label{sec:individual}

\subsubsection{The central group (B)}
\label{sec:regionB}

The center of the MAMMOTH-I Nebula was classified by \citet{cai17b} as `source B'.\footnote{The exact coordinates of source B have been verified as RA\,=\,14$^{\rm h}$41$^{\rm m}$24.456$^{\rm s}$ and dec\,=\,+40$^{\circ}$03$^{\prime}$09.45$^{\prime\prime}$ (\citealt{arr18b}; see also \citealt{cai17a}).} We adopt their terminology, and associate region B with the central group of galaxies, which is surrounded by diffuse light in the HST imaging. The CO emission in region B shows a full width at half the maximum intensity of only FWHM$_{\rm CO}$\,$\sim$\,85\,\kms\ and is observed across $\sim$4 channels ($\sim$120 \kms) in Fig.\,\ref{fig:chanmaps}. This is much narrower than the typical rotational velocity of high-$z$ galaxies \citep[e.g.,][]{ued14,cal18}. As can be seen in detail in Fig.\,\ref{fig:sourceB}, the peak of the CO emission associated with region B appears to be offset by $\sim$5\,kpc with respect to the stellar body of the central galaxies in the HST image. Interestingly, this molecular gas is co-spatial with diffuse light seen in the HST image. Furthermore, the CO emission extends across a region of $\sim$30\,kpc, and there is a hint at the $\sim$3$\sigma$ level that some of the emission stretches even further north (see the last panel of Fig.\,\ref{fig:chanmaps}). 

The unusually narrow velocity dispersion indicates that the molecular gas as traced by the CO(1-0) emission in region B is dynamically cold. The bulk of this dynamically cold gas lies just outside the galaxies in region B. It is a reasonable assumption that this dynamically cold gas has low excitation conditions, making it relatively bright in CO(1-0) compared to the warmer molecular gas that is typically found in high-$z$ galaxies using higher-$J$ lines of CO($J$,$J$-$1$). It would be interesting to study high-$J$ CO lines in search for molecular gas that is more concentrated on the stellar body of the galaxies in region B. We started an ALMA program in Cycle-6 that will address this issue.

\subsubsection{The northern galaxy (A)}
\label{sec:regionA}

\begin{figure*}
\centering
\includegraphics[width=0.95\textwidth]{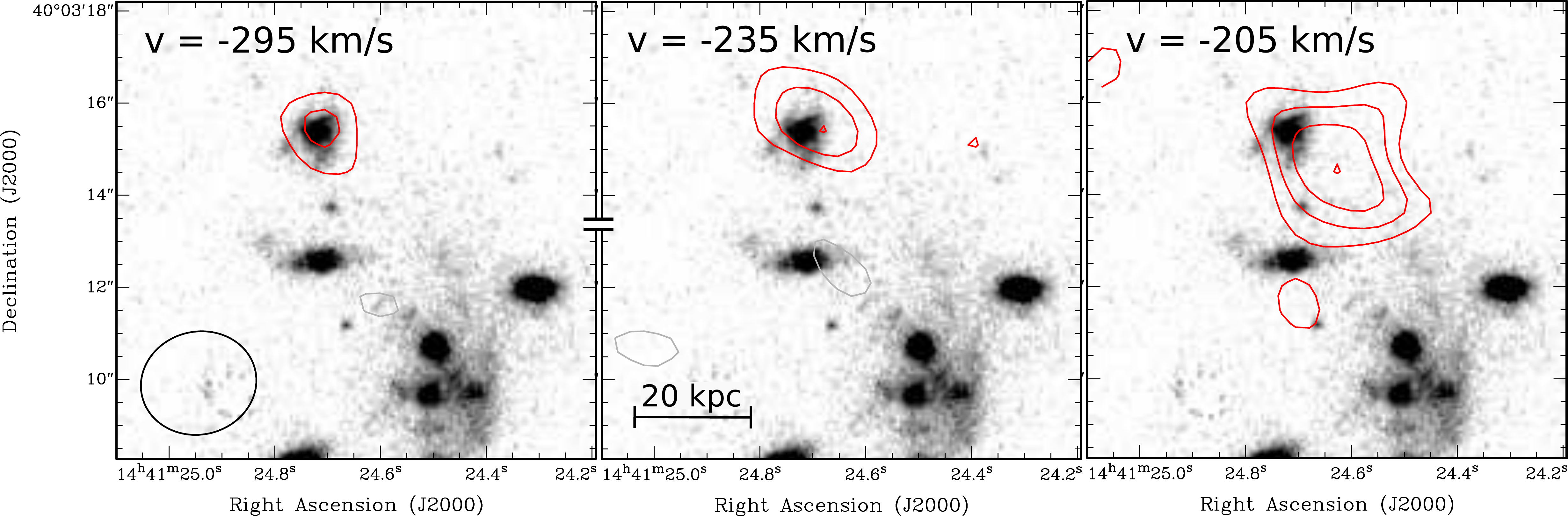}
\caption{Zoom-in of the CO(1-0) emission in region A. Contours are the same as in Fig.\ref{fig:chanmaps}.} 
\label{fig:sourceA}
\end{figure*}

About 50\,kpc north-east of region B lies a galaxy that is clearly detected in both CO(1-0) and the HST image (Fig.\,\ref{fig:sourceA}). Roughly half of the CO(1-0) emission is associated with the stellar body of the galaxy. The remaining half of the CO emission stretches outside the galaxy in what appears to be a wide tail of gas. The velocity dispersion of the gas remains approximately the same, from FWHM$_{\rm CO}$\,$\sim$\,170\,\kms\ on top of the galaxy to FWHM$_{\rm CO}$\,$\sim$\,190\,\kms\ for the extended feature. The extended emission stretches across $\sim$25\,kpc in the direction of region B, filling an area with very little light detected in the HST image. We argue that the extended feature in region A is likely either gaseous tidal debris, or an outflow.

\subsubsection{The faint western galaxy (C)}
\label{sec:regionC}

Roughly 65\,kpc west of region B we detect CO(1-0) in a galaxy that is much fainter in the rest-frame optical light than most of the other galaxies in the HST image of Fig.\,\ref{fig:mom0}. The CO is not significantly extended, and has FWHM$_{\rm CO}$\,$\sim$\,230 \kms.

\subsubsection{The southern group (D)}
\label{sec:regionD}

Approximately 100\,kpc south-west of region B, a projected group of eight galaxies is visible in the HST image of Fig.\,\ref{fig:mom0}. Figure\,\ref{fig:chanmaps} reveals tentative indications that CO(1-0) emission is associated with this group of galaxies, and that most of this emission is found between the galaxies. The brightest blob of the CO emission in region D is shown in Fig.\,\ref{fig:mom0} and its properties are given in Table \ref{tab:properties}. Tapering our data does not leave sufficient sensitivity to confirm the presence of more extended CO(1-0) emission in this region. Therefore, additional observations are required to verify and properly image the CO emission in region D.

\begin{figure*}
\centering
\includegraphics[width=0.95\textwidth]{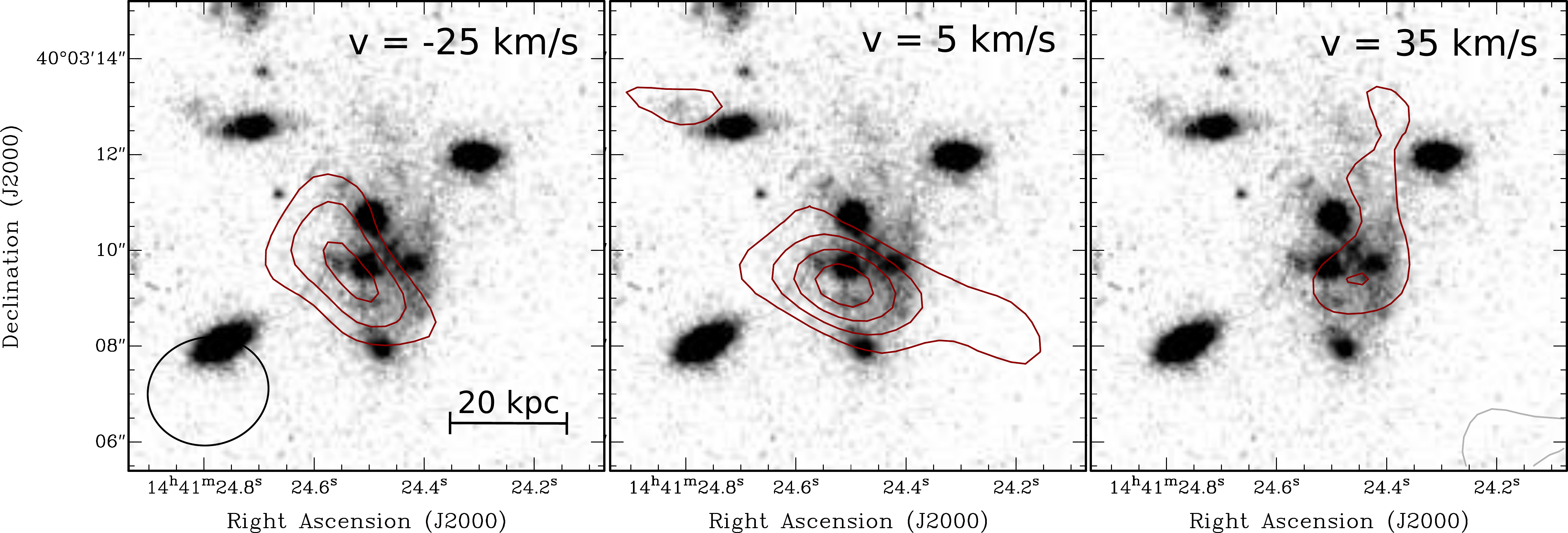}
\caption{Zoom-in of the CO(1-0) emission in region B. Contours are the same as in Fig.\ref{fig:chanmaps}.} 
\label{fig:sourceB}
\end{figure*}

\subsection{CO(1-0) in the CGM}
\label{sec:overview}

By integrating the CO(1-0) emission across the four regions described in Sect.\,\ref{sec:individual}, we derive a total CO(1-0) luminosity of $L^{\prime}_{\rm CO}$\,$\sim$\,3.8\,$\pm$\,0.8\,$\times$10$^{10}$ \Kkmspc\ in the MAMMOTH-I Nebula \citep{sol05}. Assuming a typical conversion factor for high-$z$ galaxies of $\alpha_{\rm CO}$\,$\equiv$\,M$_{\rm H2}$/$L^{\prime}_{\rm CO}$\,=\,3.6 M$_{\odot}$\,(K~\kms~pc$^{2}$)$^{-1}$ \citep[e.g.,][]{dad10,gen10}, this results in a total molecular gas mass of M$_{\rm H2}$\,$\sim$\,(1.4\,$\pm$\,0.3)\,$\times$\,10$^{11}$~M$_{\odot}$. 

A significant fraction of the CO(1-0) emission is found in the CGM between the brightest galaxies in the HST image. We note that we define the CGM as the gaseous medium outside the main stellar body of the galaxies in the HST image. As the CGM typically extends on scales of tens of kpc and the separation between the galaxies in the center of the \lya\ nebula is comparable to that, we cannot make the distinction between circum-galactic and intra-group medium with the current data, in particular in regions B and D. If we take into consideration the tail-like structure in region A, all of the emission in region B, and tentatively also the faint CO blob in region D, then we estimate $L^{\prime}_{\rm CO-CGM}$\,$\sim$\,2.2\,$\times$\,10$^{10}$ \Kkmspc\ for the cold molecular gas in the CGM. This is roughly 50\,$-$\,60\,$\%$ of the total CO(1-0) luminosity of the cold gas in MAMMOTH-I. 

We also tapered our CO(1-0) data to a beam-size of 6.4$^{\prime\prime}$\,$\times$\,5.9$^{\prime\prime}$ (53\,$\times$\,49\,kpc), resulting in an rms sensitivity of 0.1 mJy\,beam per 30 \kms\ channel. These tapered data did not recover any additional CO(1-0) emission, therefore they provide no evidence for a (diffuse) molecular gas phase that is spread across $\sim$50 kpc scales.

\section{Discussion}
\label{sec:discussion}

We presented the CO(1-0) properties across the central 150\,kpc of the enormous Ly$\alpha$ nebula of MAMMOTH-I. In various regions within the nebula, the CO emission extends across several tens of kpc. In this Section, we compare the CO emission in MAMMOTH-I with the properties of the Ly$\alpha$ nebula (Sect.\,\ref{sec:lya}), study the nature of the cold molecular CGM (Sect.\,\ref{sec:nature}), and evaluate the evolutionary state of MAMMOTH-I (Sect.\,\ref{sec:evolution}).

\subsection{Comparison of CO and Ly$\alpha$}
\label{sec:lya}

\begin{figure}
\centering
\includegraphics[width=0.47\textwidth]{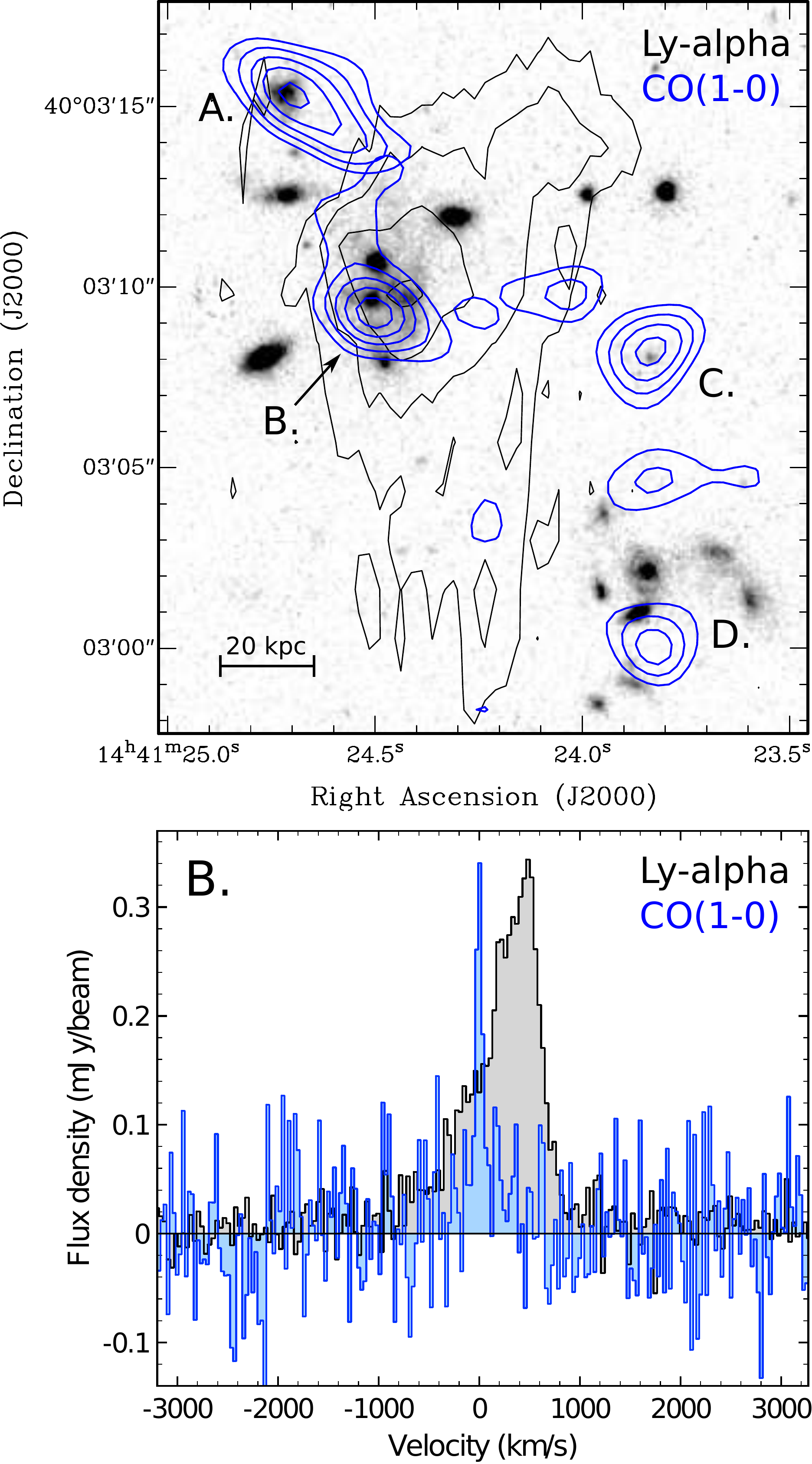}
\caption{Comparison of CO(1-0) and Ly$\alpha$ emission. {\sl Top:} CO(1-0) total intensity emission (blue contours) overlaid onto the HST/WFPC3 image from Fig.\,\ref{fig:mom0}, with added the total intensity emission of Ly$\alpha$ (black contours) obtained with the Keck Cosmic Web Imager (Li et al. in prep.). Contour levels of Ly$\alpha$ are at 5, 15, 45, 90$\%$ of the peak intensity. The emission of the resonant Ly$\alpha$ line was obtained in the range 4020$-$4040 \AA\ (-450$\,-\,$1050\,\kms). The full extent of the $\sim$440\,kpc ELAN from narrow-band imaging is shown in \citet{cai17b}. {\sl Bottom:} Overlay of the Ly$\alpha$ (black) and CO(1-0) (blue) spectra in region B. The Ly$\alpha$ line is scaled to the CO(1-0) flux density for easy comparison. Details of the Ly$\alpha$ emission will be given in Li et al (in prep).} 
\label{fig:KCWI}
\end{figure}

In Fig.\,\ref{fig:KCWI} (top) we compare the CO(1-0) emission with the Ly$\alpha$ emission detected in MAMMOTH-I. The distribution of Ly$\alpha$-emitting gas is notably different from the distribution of CO(1-0). In particular, the CO-rich regions A, C, and D are devoid of strong Ly$\alpha$ emission. Instead, the brightest patches of Ly$\alpha$ emission stretch both NW and S of region B, in areas where we find no evidence for extended CO. Extended C\,{\sc IV} also coincides with the brightest part of the \lya\ nebula (Li et al. in prep.).

In region B, the CO(1-0) emission appears to be offset from the peak of the \lya\ emission by $\sim$10 kpc ($\sim$1.2$^{\prime\prime}$). The \lya\ peak was aligned with the position of source B from \citet{cai17a} and \citet{arr18b} (see Sect. \ref{sec:regionB}). This means that there is some uncertainty in the absolute astrometry of the KCWI data. 

Figure \ref{fig:KCWI} (bottom) compares the kinematics of the CO(1-0) and Ly$\alpha$ emission in region B. Ly$\alpha$ is a resonant line, hence scattering effects, combined with geometry and absorption by neutral hydrogen, can create asymmetric and even redshifted profiles that may not reflect the true internal kinematics of the Ly$\alpha$-emitting gas \citep[e.g.][]{vil96,sta17,smi19}. Nevertheless, the Ly$\alpha$ emission in region B covers roughly 2000 \kms, while the CO(1-0) profiles shows a FWHM of only $\sim$85 \kms. This huge difference in velocity dispersion indicates that the cold molecular gas detected in CO(1-0) is not simply mixed with the warmer Ly$\alpha$-emitting gas into a single multi-phase medium. Instead, the CO emission in region B likely traces a region where cold molecular gas either formed in-situ or had time to settle within the Ly$\alpha$ nebula, as we will discuss in the next Section.

\subsection{Nature of the cold CGM}
\label{sec:nature}

In regions A and C, the CO(1-0) is associated with a galaxy. However, the CO(1-0) emission in region B, which we assume is the core of the \lya\ nebula, has a very narrow velocity dispersion (FWHM$_{\rm CO}$\,=\,85\,\kms). Moreover, this emission is offset from the main stellar body of the galaxies in the center of the nebula. Models show that the CGM in massive galaxies evolves through a complex interplay of different physical processes, including metal-enrichment, outflows, mixing, gas cooling, gas accretion, and mass transfer through galaxy interactions and mergers \citep[e.g.][]{mor06,nar15,fau16,ang17}. In the following, we discuss that a multitude of these processes must occur in the core of MAMMOTH-I.

\paragraph{Enrichment}

The presence of extended emission of both CO (Fig.\,\ref{fig:sourceB}) and C\,{\sc IV} (Li et al. in prep.) in region B indicates that widespread enrichment occurs in the MAMMOTH-I Nebula. At first glance, this does not corroborate accretion models where cold streams of relatively pristine gas directly feed massive galaxies \citep[e.g.,][]{dek09}. However, \citet{cor17} show that these streams become multi-phase and turbulent as the gas cools, and eventually mix with the ambient halo gas. It is likely that metals are deposited into this environment by winds from the galaxies or AGN in the center, or by stellar winds from stars that formed in-situ within the low gravitational potential of the CGM. This can lead to the gradual build-up of carbon, oxygen and dust \citep[e.g.][]{nar15}. In region B of MAMMOTH-I, which is close to the central AGN and starburst \citep{arr18b}, the metals and dust with which the environment is enriched serve as a catalyst for the formation of CO.

\paragraph{Gas cooling}

With enriched ambient material present in the MAMMOTH-I Nebula, it is likely that cooling processes in the core contribute to the formation of cold molecular gas. Models show that when $t_{\rm cooling}$\,$<$\,10\,$\times$\,$t_{\rm free-fall}$, cold gas clouds may continuously condense out of the galaxy's hot atmosphere, rain down on the galaxy, trigger widespread star-formation, and feed the black hole \citep{voi15,li15}. If this process occurs in MAMMOTH-I, it is likely confined to the inner region where we see the CO emission. The reason is that we do not find evidence for the presence of CO across the larger halo environment (Sect.\,\ref{sec:overview}). Moreover, the CO in region B shows velocities that are much lower than the typical velocity of clouds that form in the halo and then fall into the center, where presumably they would mix with the ambient gas to create chaotic gas motions.

Perhaps the AGN in MAMMOTH-I is an important element in cooling the CGM in the core, reminiscent of recent ALMA observations of gas cooling in brightest cluster galaxies (BCGs) at low redshifts \citep{rus16,rus17,tem18,oli19}. In these low-$z$ studies, ALMA resolved filaments of cold molecular gas with a velocity gradient and dispersion well below the velocity dispersion of the stars in the region, reaching values similar to what we observe for the CO kinematics in region B. These ALMA studies of low-$z$ BCGs, when compared with X-ray observations, also showed that the molecular gas likely originates from gas cooling in the updraft behind bubbles created by a radio-loud AGN, which buoyantly rise through the cluster atmosphere \citep{rus16,rus17,tem18}. As we will discuss in Sect.\,\ref{sec:evolution}, studies of the radio-loud Spiderweb Galaxy showed that such a scenario of AGN-driven cooling appears to be viable at high redhifts \citep[][see also models by \citealt{li14}; \citealt{mcn16}]{gul16,emo16}. However, while MAMMOTH-I contains an obscured powerful AGN \citep{arr18b}, currently there is no active radio jet detected. Given that gaseous structures with the observed low velocity dispersion would be transient on the time-scales of radio-AGN activity ($\ga$10$^{7}$ yr), additional mechanisms, such as magnetic fields, would likely be needed to prevent the cold gas from dispersing \citep[see][]{rus16}. Alternatively, for low-$z$ BCGs, galaxy interactions can cause entropy fluctuations in the hot gas that induce gas cooling \citep{oli19,van19}. This can result in narrow filaments of molecular gas, or even star formation, also in the absence of a radio-loud AGN \citep{bay02,van19}. If similar processes occur at $z$\,$\sim$\,2, then the dense group of galaxies that are offset from the peak of the narrow CO(1-0) emission in region B could potentially induce gas cooling in the core. The ELAN of MAMMOTH-I would be a prime candidate for this process to occur, but further investigation is needed.

\paragraph{Gaseous debris from galaxy interactions}

The molecular gas in region B could have been stripped from one or more galaxies through gravitational effects or ram-pressure stripping. The CO(1-0) that we observe in region B amounts to a molecular gas reservoir equivalent to that associated with the galaxy and tail-like feature in region A. However, since molecular gas is typically concentrated well within the stellar disc \citep[e.g.][]{ued14,cal18}, it is curious that we do not detect CO(1-0) emission that is clearly co-spatial with any of the galaxies detected with HST in region B. Of course, it is possible that the host galaxy itself was disrupted entirely, but such a violent event would result in gas kinematics that are much more extreme than what we observe in region B. It is more likely that the dynamically cold gas detected in region B has low excitation conditions, and is thus relatively bright in CO(1-0) compared to warmer molecular gas that may be associated with the galaxies (see Sect. \ref{sec:regionB}).

If galaxy interactions are an important factor in depositing gas in the CGM, the tidal debris must therefore have had enough time to settle and cool down in the core of the \lya\ nebula. The time needed for the molecular gas to settle down after a violent disruption event and obtain the observed narrow velocity dispersion must at least be a dynamical timescale of $t_{\rm dyn}$\,=\,R/v\,$\sim$\,10$^{8}$ yr, assuming R\,$\sim$\,20\,kpc the distance over which the molecular gas is distributed and v\,$\sim$\,200\,\kms\ the initial velocity of the gas. We note that once the gas settled into the observed narrow velocity dispersion, there may no longer be clear ways to discern the gaseous remnants of such a tidal disruption event from molecular gas that formed as a result of local gas cooling.

A way to test this scenario comes from the observation that in region B the CO(1-0) appears to follow the diffuse HST light. This could be analogue to the diffuse HST light seen across the molecular CGM of the Spiderweb Galaxy, where multi-color HST imaging revealed that the light originates from young stars that formed in-situ in the CGM \citep{hat08}. It would be interesting to obtain multi-filter HST observations also for MAMMOTH-I, to verify whether the diffuse light in the CGM represents young and blue stars formed in-situ after the gas settled and cooled down, rather than old and red tidal material from one or more recently disrupted galaxies.

\vspace{3mm}

We conclude that, while some of the CO emission in the ELAN of MAMMOTH-I is clearly associated with individual galaxies, there is also a large amount of molecular gas in the CGM that likely experienced a variety of physical processes, including local cooling. In region B, the molecular gas is dynamically cold, as shown by the narrow velocity dispersion. We therefore argue that the narrow CO(1-0) line in region B traces the systemic redshift of MAMMOTH-I at $z$\,=\,2.3116\,$\pm$\,0.0004, and that the location of the CO emission, rather than the AGN, represents the core of the potential well, and thus the true center of MAMMOTH-I.

\subsection{Evolution: an early-stage Spiderweb?}
\label{sec:evolution}

Another enormous \lya\ nebula in which widespread molecular gas has been detected is the radio-loud Spiderweb Galaxy at $z$=2.2 (\citealt{emo16,emo18}; see Sect.\,\ref{sec:intro}). MAMMOTH-I and the Spiderweb are both located in a rich proto-cluster environment at $z$\,$\sim$\,2. The CO(1-0) in region B of MAMMOTH-I appears to follow diffuse light in the restframe optical HST image. Although we lack any color-information for the HST imaging of MAMMOTH-I, this appears very similar to the CO distribution in the Spiderweb Galaxy, which follows diffuse blue light from stars that formed in-situ within the CGM \citep{hat08}.

However, there are also differences between MAMMOTH-I and the Spiderweb Galaxy. While both MAMMOTH-I and the Spiderweb Galaxy contain an obscured AGN \citep{car97,arr18b}, the Spiderweb Galaxy host a powerful radio source, but MAMMOTH-I is radio-quiet. It is not detected in the VLA FIRST (Faint Images of the Radio Sky at Twenty-Centimeters) survey down to a 3$\sigma$ limit of 0.42 mJy~beam$^{-1}$, which corresponds to $P_{\rm 1.4\,GHz}$\,$\le$\,1.4\,$\sim$\,10$^{24}$~W\,Hz$^{-1}$. In the Spiderweb Galaxy, both H$_{2}$O and enhanced CO(1-0) emission are found along the radio jet, indicating that energy injected into the CGM by the propagating radio jet can cause local thermal instability that leads to gas cooling \citep[][see also models by \citealt{li14}]{gul16,emo16}. This process is currently absent in MAMMOTH-I, but given that an AGN is expected to be radio-loud $\sim$10$\%$ of the time, a phase of more rapid gas cooling could occur during subsequent episodes in the evolution of MAMMOTH-I. In addition, compared to what we observe in MAMMOTH-I, in the Spiderweb nebula the diffuse light appears to be more widespread \citep{hat08} and the concentration of galaxies is denser \citep{mil06}. Also, the CO within the ELAN of MAMMOTH-I is partially associated with individual galaxies (regions A and C), while in the Spiderweb Galaxy the cold molecular gas is more settled into a single large gas reservoir in the CGM, which has a significantly lower velocity dispersion than the galaxies \citep{emo16}. 

The above similarities and differences raise the intriguing possibility that MAMMOTH-I is a system much like the Spiderweb Galaxy, but in an earlier stage of evolution.

\section{Conclusions}
\label{sec:conclusions}

We used the VLA in its most compact D-configuration to trace CO(1-0) emission of cold molecular gas in the Enormous Ly$\alpha$ Nebula of MAMMOTH-I ($z$\,=\,2.3). We detected CO in four regions that are associated with either galaxies or groups of galaxies that occupy the inner 150 kpc of the nebula. We derive the following main conclusions:\\
\vspace{-2mm}\\
$\bullet$ The total mass of molecular gas that we detect in MAMMOTH-I is M$_{\rm H2}$\,=\,1.4\,($\alpha_{\rm CO}$/3.6)\,$\times$\,10$^{11}$ M$_{\odot}$.\\
\vspace{-2mm}\\
$\bullet$ Roughly 50$\%$ of the CO emission is found at various places in the CGM, stretching across scales of up to $\sim$30 kpc. The other $\sim$50$\%$ of the CO is associated with the main stellar body of two galaxies, one which shows a wide gaseous tidal-tail or outflow, the other one is very faint in the optical rest-frame.\\
\vspace{-2mm}\\
$\bullet$ In the center, the CO(1-0) in the CGM has an unusually narrow velocity dispersion (FWHM\,$\sim$\,85 \kms) and follows diffuse light seen in HST imaging. This is dynamically cold gas that likely originated from the cooling of an enriched multi-phase medium in the core of the potential well of MAMMOTH-I.\\
\vspace{-2mm}\\
In Sect.\,\ref{sec:evolution} we speculated that the CO-rich Ly$\alpha$ Nebula of MAMMOTH-I bears a resemblance to the enigmatic Spiderweb Galaxy, but perhaps represents an earlier stage in evolution. To trace the molecular CGM on large scales around high-$z$ massive galaxies and proto-clusters requires sensitive low-surface-brightness observations using very compact millimeter interferometers \citep{emo18b}. The number of observations that have reached the surface-brightness levels needed to detect very extended CO emission is still limited. Dedicated millimeter observations in search for widespread molecular gas in the CGM of other Enormous Ly$\alpha$ Nebulae may better reveal how these massive structures evolve into the most massive cluster galaxies in the Universe.


\acknowledgments
BE wishes to thank Gustaaf van Moorsel for technical help with the VLA observations, and for being a great colleague and valued collaborator over the years. The team also thanks William Mathews for interesting discussions on this work. The National Radio Astronomy Observatory is a facility of the National Science Foundation operated under cooperative agreement by Associated Universities, Inc. ZC acknowledges the supports provided by NASA through the Hubble Fellowship grant HST-HF2-51370 awarded by the Space Telescope Science Institute, which is operated by the Association of Universities for Research in Astronomy, Inc., for NASA, under contract NAS 5-26555. JXP acknowledges support from the National Science Foundation (NSF) grant AST-1412981. Based on observations made with the NASA/ESA Hubble Space Telescope, obtained from the data archive at the Space Telescope Science Institute. STScI is operated by the Association of Universities for Research in Astronomy, Inc. under NASA contract NAS 5-26555. Some of the data presented herein were obtained at the W. M. Keck Observatory, which is operated as a scientific partnership among the California Institute of Technology, the University of California and the National Aeronautics and Space Administration. The Observatory was made possible by the generous financial support of the W. M. Keck Foundation. This work has made use of data from the European Space Agency (ESA) mission {\it Gaia} (\url{https://www.cosmos.esa.int/gaia}), processed by the {\it Gaia} Data Processing and Analysis Consortium (DPAC, \url{https://www.cosmos.esa.int/web/gaia/dpac/consortium}). Funding for the DPAC has been provided by national institutions, in particular the institutions participating in the {\it Gaia} Multilateral Agreement.\\

%

\vspace{5mm}
\facilities{VLA, HST(WFC3), Keck(KCWI), GAIA}


\software{CASA \citep{mcm07}}



\end{document}